# Estimating Asset Class Health Indices in Power Systems

Ming Dong
Grid Analytics
*Alberta Electric System Operator*

*Abstract*— Power systems have widely adopted the concept of health index to describe asset health statuses and choose proper asset management actions. The existing application and research works have been focused on determining the current or near-future asset health index based on the current condition data. For preventative asset management, it is highly desirable to estimate asset health indices, especially for asset classes in which the assets share similar electrical and/or mechanical characteristics. This important problem has not been sufficiently addressed. This paper proposes a sequence learning based method to estimate health indices for power asset classes. A comprehensive data-driven method based on sequence learning is presented and solid tests are conducted based on real utility data. The proposed method revealed superior performance with comparison to other Estimation methods.

## I. Introduction

POWER systems are asset intensive. Today, many utilities have been requested by regulators to develop cost-effective asset management and maintenance strategies to reduce overall cost while still maintaining acceptable system reliability [1]. To achieve this goal, they have adopted proactive asset inspection programs and asset condition data management systems to support data based decision making. One effective and straightforward method is to use an asset health indexing system [2-3], in which several asset health index levels (such as 1 to 5) are assigned to the assets based on their current health condition data. For different index levels, different asset management actions can be taken accordingly. In [3], an international energy research institute surveyed 21 electric utility companies in US, Canada, Australia and Israel and summarized typical asset management actions against health index levels for typical power assets. An asset with a high health index (indicating very healthy status) needs minimum attention and may be suitable for a run-to-fail strategy; an asset with a low health index (indicating very unhealthy status) should be preventatively replaced to minimize potential system losses; an asset with a medium health index should be inspected and maintained at a frequency higher than that in the initial stage. Typical health index definitions and actions are shown in Table I.

TABLE I: TYPICAL HEALTH INDEX DEFINITION AND ASSET MANAGEMENT ACTIONS

| Health Index | Definition | Asset Management Action |
|---|---|---|
| 5 | In "as new" condition | Minor Maintenance |
| 4 | Has some minor problems or evidence of aging | Normal Maintenance |
| 3 | Has many minor problems or a major problem. Without intervention, problem(s) would accelerate aging rate | Increase asset inspection and maintenance frequency |
| 2 | Has many serious problems. Without intervention, problems may cause asset failure | Start planning process for asset replacement or rehabilitation |
| 1 | Asset has deteriorated to the stage where failure is imminent | Asset has reached its end-of-life and should be replaced immediately. |

In general, there are two kinds of methods to determine the asset health indices similar to those listed above:
- Rely on human experts such as asset inspectors or engineers to analyze asset conditions and assign a proper health index to an asset on an individual basis. This method can be time consuming but may be more accurate.
- Calculate health indices based on individual asset condition attributes. Weighting factors are used to combine different condition scores into one overall health index for an asset in [3-6]. However, the weighting factors are often difficult to determine and can vary significantly among different utilities and/or areas due to weather conditions, operating conditions and maintenance practices. Instead, [7] developed machine learning models from utility data to calculate the health indices of power transformers.

The existing methods are generally based on the idea of mapping current asset condition data to a current or near-future asset health index. To our best knowledge, there is no or little literature for estimating an asset health index for a horizon (such as in a few years), although this is highly desirable from a preventative asset management planning perspective, especially for asset classes.

A power asset class is defined as a group of equipment that has the same function and shares similar electrical and/or mechanical characteristics. In a real power system, majority of assets belong to asset classes. Common examples are certain types of cables, conductors and poles, secondary transformers in a service area. Not like costly standalone assets such as substation power transformers, the assets in asset classes are often large in quantity and hence do not have costly online monitoring systems that can continuously monitor their health and react to unfavorable conditions in a short timeframe. What is available for asset classes is the asset condition data recorded through inspections programs which are often periodically conducted. Currently, utilities Estimate asset failure rates for asset management [8-11]. Different from the failure rates that are based on statistics of failures, health indices are made from multiple condition attributes and directly mapped to different sets of asset management actions by asset type. Therefore, it would be extremely beneficial to estimate health indices for power asset classes.

This paper proposes a novel sequence learning based method to Estimate asset health indices in asset classes. Recently, sequence learning has been successfully applied to various forecasting problems such as the forecast of power demand, stock price and weather conditions [12-15]. A major advantage of sequence learning is that it can systematically utilize historical data and capture the pattern of feature variations over time. This method requires sufficient historical data which is very suitable for asset classes because the assets in an asset class are essentially the same and a considerable amount of historical condition data is available. The main contributions of this paper include the following:

- A general framework for Estimateing health indices for power asset classes is presented;
- A unique model architecture is establshed to convert a health index Estimationproblem into a temporal multi-class classification problem;
- The sequence learning model, which can effectively extract and utilize the information embedded in the inspection history, is thoroughly tested. The results indicate that this kind of model can significantly improve the Estimationaccuracy.

The paper is organized as follows. The current utility practice for gathering and storing different types of asset condition data is described in Section II. The feature engineering techniques of converting raw asset condition data into features suitable for sequence learning are explained in Section III. Long short-term memory network based sequence learning theory along with proper model architecture and training dataset structure is explained in Section IV. The proposed method is validated using a Canadian utility company's wood pole and underground cable asset class datasets and compared to other Estimationmethods in Section V, followed by discussions and conclusions in Section VI.

## II. HISTORICAL ASSET CONDITION DATA

Many utility companies have realized the value embedded in data and started to use data to support complex decision making. Specifically for asset class management, two main efforts have been made in the industry:

- Utility companies established sophisticated inspection programs for asset classes to gather asset condition data [1-3]. For example, many Canadian utility companies have wood pole, underground cable and service transformer and other inspection programs to inspect assets every a few years.
- The results of the above inspection, i.e. the asset condition data are recorded and stored in the Computerized Maintenance Management Systems (CMMS) to support advanced data analytics [16].

The above industry effort has made the proposed data-driven method possible for health index Estimation. For illustration purpose, an example of asset condition dataset is given in Table II. It should be noted that the data stored in the CMMS system may not directly look like this. Through proper data manipulation, the data can be organized in a way similar to the given format for analytics purpose.

TABLE II: A HISTORICAL ASSET CONDITION DATASET EXAMPLE FOR A POWER ASSET CLASS

| Asset ID | Inspection Time | Condition Attribute | | | | | Asset Service Age |
|---|---|---|---|---|---|---|---|
| | | $C_1^n$ | $C_2^n$ | $C_3^n$ | $C_1^o$ | $C_1^u$ | |
| 00001 | 2015 | 26 | -2.35 | 198 | Medium | Type 1 | 16 |
| 00001 | 2012 | 20 | -1.37 | 197 | Medium | Type 1 | 13 |
| 00001 | 2009 | 5 | 0.42 | 201 | Moderate | Type 1 | 10 |
| 00002 | 2015 | 37 | 0.78 | 143 | Severe | Type 2 | 18 |
| 00002 | 2012 | 32 | 1.51 | 143 | Medium | Type 2 | 15 |
| 00002 | 2009 | 22 | 1.69 | 146 | Medium | Type 1 | 12 |
| … | | … | … | … | … | … | |

Table II contains all available assets that belong to the same power asset class, labeled by their unique asset ID. Each asset has multiple asset condition rows recorded at different inspection times. In addition to asset service age at the time of inspection, multiple condition attributes are recorded. For example, for a SF6 pad-mounted switchgear, the SF6 pressure gauge reading, condition of enclosure, condition of terminations, condition of blades, condition of pad, Infrared scan results can be all gathered in one inspection [3]. In general, there are three types of condition data in a condition dataset: numerical, ordered categorical data and unordered categorical data. It is important to understand the differences between these types because different feature engineering techniques have to be applied to them at the later feature engineering stage for machine learning purpose:
- Numerical data: this type of data is continuous real numbers ($C_1^n$, $C_2^n$ and $C_3^n$ in Table II). Examples are transformer bushing leak current, SF6 switchgear pressure gauge reading, pole leaning angle, cable partial discharge inception and extinction voltages, service loading and etc.
- Ordered categorical data: this type of data is not based on accurate measurement but also widely exists in the inspection results. For example, pole surface damage can be rated to low, medium or high ($C_1^o$ in Table II); enclosure rusting condition can be rated to 1 to 5. This type of data often involves engineering judgment from inspectors based on their inspection experience on the same asset class.
- Unordered categorical data: this type of data is often used to describe the operating conditions of assets which may also affect the result of health index assignment ($C_1^u$ in Table II). It is not ordered because it does not directly indicate the asset health is better or worse between different values. However, it is a parameter that should be considered in the health Estimateion. For example, the same type of poles may or may not have overhead transformers mounted to them. Different weight loads have different impact on the life of poles; the same type of cables can be used to supply different types of loads. Certain industrial load may contain more harmonic content and is more likely to cause cable overheating issues [17].

## III. FEATURE ENGINEERING

This section explains the necessary feature engineering techniques to convert the raw condition data discussed in Section II to features that can be processed by the subsequent sequence learning model. A few problems need to be solved in the feature engineering step: categorical data should be converted to numerical features; numerical condition data should be normalized; the dimension of multivariate condition features can be reduced so that the future learning process can become more effective and accurate.

*A. Conversion of Ordered Categorical Asset Condition Data to Numerical Features*

Categorical data cannot be directly processed by machine learning models. According to [18], ordered categorical condition data such as low, medium, high can be converted to numerical features by using the equation below:

$$x = \frac{i - 1/2}{N}, i = 1,2 \dots N \quad (1)$$

where $N$ is the total number of ratings, $i$ is the order of a rating.

For example, for a low-medium-high rating system, the low rating can be converted to 0.17; the medium rating can be converted to 0.5; the high rating can be converted to 0.83.

*B. Conversion of Unordered Categorical Asset Condition Data to Numerical Features*

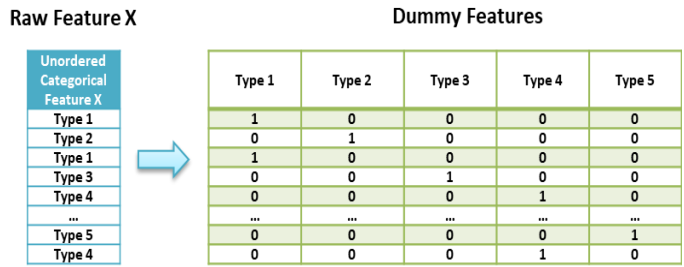

Fig.1. Conversion of unordered categorical asset condition data to dummy features

Unlike ordered categorical data, unordered categorical data cannot be directly converted to numerical features but a feature engineering technique called "Dummy Features" can be used for indirect conversion [19]. As shown in Fig.1, one unordered categorical feature with $N$ types to select from can be converted to $N$ dummy features. Each dummy feature is either 0 or 1. For example, being type 3 in a 5-type system can be converted to feature vector (0,0,1,0,0); being type 5 can be converted to vector feature (0,0,0,0,1).

*C. Feature Normalization*

All numerical features should be normalized to a fixed numeric range such as [0,1]. This is because the raw condition data have different units and the difference of feature magnitudes can be quite large. There are many ways to normalize features such as the Min-Max normalization [19]:

$$x_{norm} = \frac{x_{raw} - Min}{Max - Min} \quad (2)$$

For a given dataset, $x_{raw}$ is a raw feature value; $Max$ is the maximum value observed in that feature column; $Min$ is the minimum value observed in that feature column.

*D. Feature Dimension Reduction*

Table II contains many asset condition features. These features describe different conditional aspects but can be highly correlated. For example, a piece of aging equipment may have multiple conditions deteriorating all together. This implies the asset conditions could be compressed. In other words, reduction of feature dimension is possible, especially for the numerical ones (including the converted ones). The benefit of feature reduction is that most machine learning models are susceptible to so called dimension disaster problem [19]. When the ratio between the number of training records and the number of features is too low, the model may not get trained effectively and can run into over-fitting problem. When such a problem occurs, the model may perform well on the test set but may fail to generalize on new datasets. This problem is especially prominent if historical asset condition data is limited.

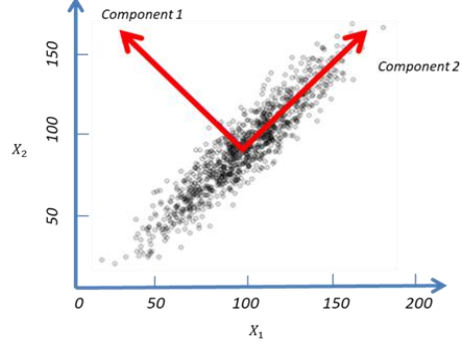

Fig.2. A PCA Example

There are many ways to reduce feature dimension. One way is to quantify the importance of different features and directly remove the less important ones [20]. Another way is to apply special feature transformation algorithms such as Principal Component Analysis (PCA) [19]. PCA is as an orthogonal linear matrix transformation that projects the data points with multiple features to a new coordinate system. In this new coordinate system, coordinates are independent to each other and on each coordinate, the magnitude of data variance is different. The coordinates which show large data variances are deemed as important coordinates because they contain more information embedded in the data. They can be kept and become a set of principal components. Projecting data points onto these important coordinates forms up a new set of features with reduced dimension. As shown in Fig.2, $X_1$ and $X_2$ are original coordinates for the presented data points. PCA establishes a different set of orthogonal coordinates (Component 1- Component 2) and in this new coordinate system, it is obvious that Component 2 represents more variance of data points than Component 1. If only one dimension is allowed to keep, Component 2 can be selected as the principal component.

Mathematically, the above transformation is defined by the following equation:

$$X^T X = PDP^{-1} \qquad (3)$$

where $X$ is the pre-processed data matrix with $n$ feature columns from the original data matrix. In $X$, the values in each feature column are normalized and subtracted by the column mean. This way, $X^T X$ essentially becomes the covariance matrix of $X$; $X^T X$ is then Eigen-decomposed; $P$ is a *n*-by-*n* matrix in which the columns are eigenvectors of $X^T X$; $D$ is a diagonal matrix with eigenvalues $\lambda_i$ on its diagonal and zeros everywhere else.

After this step, the transformed data matrix is calculated by:

$$T = XP \qquad (4)$$

In data matrix $T$, each column has the projected values on a new coordinate (component). Important features can be selected with respect to the eigenvalues obtained in the diagonal matrix *D*. This can be measured by using the metric Proportion of Variance Explained (*PVE*). *PVE* of *t* features indicates the amount of information (variance) that is kept in such features and is mathematically calculated in (5). *j* principal components can be selected accordingly to ensure their *PVE* is above a certain acceptable threshold such as 90%.

$$PVE = \frac{\sum_{j=1}^{t} \lambda_j}{\sum_{i=1}^{n} \lambda_i} \qquad (5)$$

IV. UTILITY VALIDATION AND CASE STUDIES

The proposed method was applied to a major utility company in Western Canada for the estimation of its distribution wood pole and underground cable populations in a service area. The datasets, validation results and case studies are discussed as follows.

*A. Dataset Description*

The wood pole dataset contains 3000 45-foot western red cedar wood poles in the area and the following condition attributes are gathered historically through the inspection program at a 10-year interval in 1998 and 2008.
- Shell thickness: this is a numerical condition attribute. During inspection, inspectors drill small holes around the pole bottom to measure the shell thickness. It reflects the internal rot condition due to infestation and moisture [25]. Three measurements drilled at 120 apart are taken.
- Ground line pole circumference: this is a numerical condition attribute. Reduced circumference at ground line is an indication of presence of external rot [25].
- Surface conditions: this is an ordered categorical attribute. Pole inspectors rate the surface conditions among poor, medium and good based on the presence of cracks, burns, cuts and other types of surface defects.
- Wood pecker holes: this is a binary unordered categorical attribute indicating the presence of wood pecker holes along the body of the pole.
- Carrying Transformer: this is an unordered categorical attribute with three statuses: carrying no transformer, a single-phase transformer or a three-phase transformer bank.

The cable dataset contains 2500 single-phase 13.8kV XLPE cable segments and the following condition attributes are gathered historically through the inspection program at a 5-year interval in 2003, 2008 and 2013:
- Partial discharge intensity: this is a numerical condition attribute. Water trees and voids in the insulation and other defects can lead to partial discharge activities inside the cable [26]. Partial discharge intensity can be measured in pico-coulombs (pC) during inspection test [27].
- Neutral corrosion condition: This is a numerical condition attribute. Corroded neutral may lead to a cable failure. Neutral corrosion condition is tested using time domain reflectometer (TDR) during inspection [28].
- Visual condition: This is an ordered categorical condition attribute. Inspectors look for cable discoloration, surface cracks, and surface contamination [28]. A health rating among poor, medium and good is assigned.
- Average loading condition: This is a numerical condition attribute in Amps estimated from power flow software.

In 2018, health index $H_1$ to $H_5$ of all assets in the above two datasets were assigned by asset inspectors 5 years after their last pole inspection in 2013 and 3 years after their last cable inspection in 2015. The health index composition of the two datasets is shown in Fig.6.

*B. Estimation Results for Wood Poles and Cables*

To test the proposed method, both dataset was split into training set and test set based on an 80:20 ratio. The test set was used to evaluate the estimation performance. For classification tasks, Precision and Recall are widely applied evaluation metrics. Precision is the percentage of relevant instances among the retrieved instances while recall is the percentage of the relevant instances that were actually retrieved.

For a specific health index level, its Precision index and the Recall index can be calculated based on the counts of True Positive (TP), False Positive (FP) and False Negative (FN) as below [29]:

$$\begin{cases} P = \frac{TP}{TP+FP} \\ R = \frac{TP}{TP+FN} \end{cases} \quad (17)$$

For multi-class classification tasks such as the discussed problem, Macro-precision (MP) and Macro-recall (MR) can be used to describe the overall classification performance and they are the average of precision and recall values for each class, i.e. health index level:

$$\begin{cases} MP = \frac{1}{N}\sum_{i=1}^{N} P_i \\ MR = \frac{1}{N}\sum_{i=1}^{N} R_i \end{cases} \quad (18)$$

where *N* is the number of health index levels, in this case 5.

Two sequence learning models were established for the two datasets respectively. The pole model contains 2 LSTM units while cable model contains 3 LSTM units both corresponding to their inspection years. The pole model has an input layer with 6 neurons (after feature engineering) and two hidden layers with 10 neurons; the cable model has an input layer with 4 neurons and two hidden layers with 8 neurons. Both models have a 5-neuron output layer and a softmax layer corresponding to $H_1$ to $H_5$ as illustrated in Fig.5. The ReLU activation functions are used in both models. The above hyper-parameters were optimized by using the grid search method. These two models were trained with their training sets and tested on the pre-split test sets. Their test results by Precision, Recall, MP and MR are given in Table IV.

Table IV: EstimationEvaluation

| Asset | Health Index | $H_1$ | $H_2$ | $H_3$ | $H_4$ | $H_5$ | Overall |
|---|---|---|---|---|---|---|---|
| Pole | Precision/MP | 0.83 | 0.73 | 0.86 | 0.88 | 0.84 | 0.83 |
|  | Recall/MR | 0.76 | 0.79 | 0.90 | 0.81 | 0.85 | 0.82 |
| Cable | Precision/MP | 0.94 | 0.86 | 0.81 | 0.80 | 0.92 | 0.87 |
|  | Recall/MR | 0.87 | 0.84 | 0.93 | 0.88 | 0.82 | 0.87 |

As a multi-class classification problem, the results of both assets are quite accurate. This indicates the proposed method can comprehensively analyze the asset condition data obtained from previous inspections and accurately Estimate asset health indices in a few years. Although the models were trained based on 2018 health indexing results, they can be continuously applied to Estimate health indices in future years once new condition data becomes available through future inspection. Furthermore, although trained and tested based on fixed datasets, the established models can be used for health index estimation of other poles and cables belonging to the same asset classes in the area. Such estimations very useful for preventative asset management.

*C. Comparison with using Health Index Data in Latest Inspection Year Only*

To reveal the advantage of using condition data recorded at different inspection years for Estimation, the regular Feed-forward Neural Network (FNN) model is adopted in comparison. Different from the LSTM model, it only takes in the condition data obtained from the last inspection year as the input. The same feature engineering step and softmax layer is used. The results are shown in Fig.7.

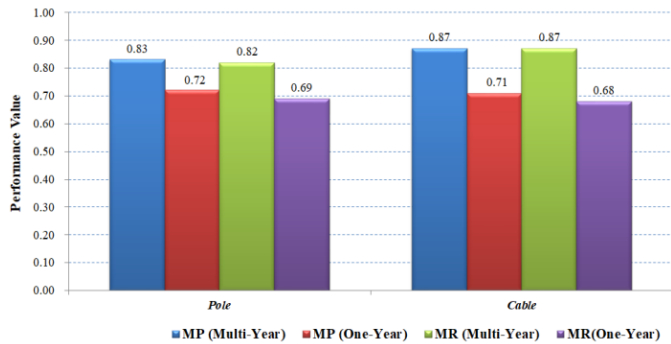

Fig.7. Comparison with using data from latest inspection year only

Both MP and MR are significantly reduced if only one year inspection data is used. This indicates the condition variation of previous years plays a key role in Estimation because assets with similar conditions can deteriorate at different paces.

V. DISCUSSIONS AND CONCLUSIONS

This paper discussed an important but overlooked problem for utility preventative asset management - health index estimation for power asset classes. The importance of this problem is due to the following two reasons:
- In the current utility practice, asset health indices are determined only for the current or near-future state. To achieve preventative asset management, it is necessary for utilities to Estimate asset health indices so that proper budget/resources can be planned and preventative actions can be taken to prevent asset failures from actually happening.
- It is especially necessary to conduct a health index estimation for assets in asset classes. This is because they are not equipped with costly online monitoring devices that critical standalone assets could have. As a result, their failures cannot be timely detected. Power asset classes commonly rely on periodical inspection programs and hence a estimation of their health is extremely beneficial.

To address the problem, a comprehensive data-driven method based on sequence learning model LSTM is proposed in this paper. The detailed validation is completed using real utility datasets for two power asset classes. The satisfactory estimation accuracy has been achieved. The proposed method is also compared to the method of only using one-time inspection data and the methods of using various other popular estimation models. The comparison reveals the superior performance of the proposed method in the health index estimation of power asset classes.

REFERENCES

[1] Asset Management-Overview, Principles and Terminology, ISO 55000- 2014.
[2] Canadian Electricity Association, "Asset Health Indices: A Utility Industry Necessity", Manitoba Public Utilities Board, MN, Canada,[Online].Available: http://www.pubmanitoba.ca/v1/exhibits/mh_gra_2015/coalition-10-3.pdf

[3] The Centre for Energy Advancement through Technological Innovation, "Distribution System Health Indices",Canada,[Online].Available: https://www.ceati.com/projects/publications/publication-details/?pid=50%2F118

[4] J. Haema and R. Phadungthin, "Condition assessment of the health index for power transformer," 2012 Power Engineering and Automation Conference, Wuhan, 2012, pp. 1-4.

[5] J. P. Lata, D. P. Chacón-Troya and R. D. Medina, "Improved tool for power transformer health index analysis," 2017 IEEE XXIV International Conference on Electronics, Electrical Engineering and Computing (INTERCON), Cusco, 2017, pp. 1-4.

[6] A. Said and R. A. Ghunem, "A Techno-Economic Framework for Replacing Aged XLPE Cables in the Distribution Network," IEEE Trans. on Power Delivery, doi: 10.1109/TPWRD.2020.2967141 (early access).

[7] K. Benhmed, A. Mooman, A. Younes, K. Shaban and A. El-Hag, "Feature selection for effective health Index diagnoses of power transformers," IEEE Trans. on Power Delivery, vol. 33, no. 6, pp. 3223-3226, Dec. 2018.

[8] W. Li, "Incorporating aging failures in power system reliability evaluation," IEEE Trans. Power Systems, vol. 17, no. 3, pp. 918-923, Aug. 2002.

[9] M. Dong and A. B. Nassif, "Combining modified weibull distribution models for power system reliability forecast", IEEE Trans. on Power Systems, vol. 34, no. 2, pp. 1610-1619, March 2019.

[10] J. Zhong, W. Li, R. Billinton and J. Yu, "Incorporating a Condition Monitoring Based Aging Failure Model of a Circuit Breaker in Substation Reliability Assessment," IEEE Trans. on Power Systems, vol. 30, no. 6, pp. 3407-3415, Nov. 2015.

[11] M. Buhari, V. Levi and S. K. E. Awadallah, "Modelling of Ageing Distribution Cable for Replacement Planning," IEEE Trans. on Power Systems, vol. 31, no. 5, pp. 3996-4004, Sept. 2016.

[12] R. Sun and C. L. Giles, "Sequence learning: from recognition and Estimationto sequential decision making," IEEE Intelligent Systems, vol. 16, no. 4, pp. 67-70, July-Aug. 2001.

[13] S. Bouktif, A. Fiaz, A. Ouni and M. A. Serhani,, "Optimal deep learning lstm model for electric load forecasting using feature selection and genetic algorithm: Comparison with machine learning approaches," Energies, vol.11, no.7,pp.1636, 2018.

[14] S. Liu, G. Liao and Y. Ding, "Stock transaction Estimationmodeling and analysis based on LSTM," 13th IEEE Conference on Industrial Electronics and Applications (ICIEA), Wuhan, 2018, pp. 2787-2790.

[15] M.A.Zaytar and C. El Amrani, "Sequence to sequence weather forecasting with long short term memory recurrent neural networks", Int J Comput Appl, vol.143, no. 11, 2016.

[16] Z. Huo, Z. Zhang, "CMMS and its application in power systems," Systems, 2003 IEEE Intl. Conf. Man and Cybernetics, DC, 2003, vol.5, pp. 4607-4612.

[17] A. Hiranandani, "Calculation of cable ampacities including the effects of harmonics," IEEE Industry Applications Magazine, vol. 4, no. 2, pp. 42-51, March-April 1998.

[18] A.Ahmad and L.Dey, "A k-mean clustering algorithm for mixed numeric and categorical data," Data & Knowledge Engineering, vol.63, no.2, pp.503-527, Apr.2007.

[19] I.H. Witten, E. Frank, M.A. Hall, and C.J. Pal, Data Mining: Practical Machine Learning Tools and Techniques, Morgan Kaufmann, 2016.

[20] B. H. Menze, B. M. Kelm, R. Masuch, U. Himmelreich, P. Bachert, W. Petrich and F. A. Hamprecht, "A comparison of random forest and its Gini importance with standard chemometric methods for the feature selection and classification of spectral data," BMC Bioinformatics, vol.10, no.1, pp.213, Dec.2009.

[21] S. Hochreiter and J. Schmidhuber, "Long short-term memory," Neural Computation, vol.9, pp.1735-1780, 1997.

[22] J. F. Kolen and S. C. Kremer, "Gradient flow in recurrent nets: the difficulty of learning dependencies," A Field Guide to Dynamical Recurrent Networks , IEEE, 2001.

[23] R. A. Dunne, and A. C. Norm, "On the pairing of the softmax activation and cross-entropy penalty functions and the derivation of the softmax activation function," Proc. 8th Aust. Conf. on the Neural Networks, Melbourne, vol. 181, p. 185. 1997.

[24] S. Mannor, D. Peleg, and R. Rubinstein, "The cross entropy method for classification," Proceedings of the 22nd International Conference on Machine Learning, pp. 561-568. 2005.

[25] J. J. Morrell, "Wood Pole Maintenance Manel: 2012 Edition", Forest Research Laboratory, Oregon State University, 2012.

[26] M.Villaran and R.Lofaro, "Condition Monitoring of Cables", Brookhaven National Laboratory, NY, USA, [Online].Available: https://www.bnl.gov/isd/documents/70782.pdf

[27] Concentric Neutral Cables Rated 5 Through 46 KV, ICEA Standard S-94-649 2013 Edition.

[28] G.J. Bertini, "Neutral Corrosion-Significance, Causes & Mitigation", Novinium Inc., WA, USA. [Online].Available: https://www.novinium.com/wp-content/uploads/2015/05/Neutral_Corrosion-Significance-Causes-Mitigation.pdf

[29] D. M. Powers, "Evaluation: from precision, recall and F-measure to ROC, informedness, markedness and correlation," International Journal of Machine Learning Technology, vol.2,no.1, pp.37-63, 2011.